\documentstyle[preprint,aps,tighten]{revtex}
\begin{document}

\title{$S_{\rm eff}(E)$ and the $^7$Be$(p,\gamma)^8$B reaction}
\author{B.~K.~Jennings and S.~Karataglidis}
\address{TRIUMF, 4004 Wesbrook Mall, Vancouver, British Columbia, Canada, 
V6T 2A3}
\date{\today}
\maketitle
\begin{abstract}
We explore approximations to the effective $S$ factor for the
$^7$Be$(p,\gamma)^8$B reaction using different approximation for the
integral over the Gamov peak. In the temperature range of
interest for solar neutrino production, $S_{\rm eff}(E)$ may be
determined to within 0.5\% from $S(20)$ with no direct information on
the derivatives of $S(E)$.
\end{abstract}

The $^7$Be($p,\gamma$)$^8$B reaction, at energies of approximately
20~keV, plays an important role in the production of solar
neutrinos \cite{bahcall}. The subsequent decay of the $^8$B is the
source of the high energy neutrinos to which many solar neutrino
detectors are sensitive. The cross section for this reaction is
conventionally expressed in terms of the $S$ factor which is defined
in terms of the cross section, $\sigma$, by
\begin{equation}
S(E) = \sigma(E) E \exp{\left[ 2\pi\eta(E) \right]}\;,
\label{eq-define}
\end{equation}
where $\eta(E) = Z_1 Z_2 \alpha \sqrt{\mu c^2/2E}$ is the Sommerfeld
parameter, $\alpha$ is the fine structure constant, and $\mu$ is the
reduced mass. The reaction rate per particle pair is \cite{bahcall2}
\begin{equation}
\left\langle \sigma\nu \right\rangle = \left( \frac{8}{\pi\mu} \right)^2
\frac{1}{(kT)^{3/2}}\int_0^\infty dE\; S(E)
\exp\left[-\frac{E}{kT} - \frac{b}{E^{1/2}}\right] 
\label{eq-reaction} 
\end{equation}
where $T$ is the temperature, $k$ is the Boltzmann factor and
$b=(2\mu)^{1/2} \pi e^2 Z_1 Z_2/\hbar$ comes from the Coulomb penetration
factor. In order to study the approximations to the integral in
Eq.~(\ref{eq-reaction}) it is convenient to introduce an effective $S$
factor defined by \cite{bahcall2}
\begin{equation}
S_{\rm eff}(E) = \exp{[\tau]} E_0 \sqrt{\frac{\tau}{4\pi}} \int_0^\infty
dE \; S(E) \exp{ \left[ -\frac{E}{kT} - \frac{b}{E^{1/2}} \right] }
\label{eq-seff}
\end{equation}
where $\tau = 3 E_0 /kT$ and $E_0 = (bkT/2)^{2/3}$ is the location at
which the function $\exp{ \left[ -\frac{E}{kT} - \frac{b}{E^{1/2}}
\right] }$ peaks. Since the integrand is strongly peaked the integral
may be done by the saddle point method \cite{bahcall2}. To first order
in $1/\tau$ the result is \cite{bahcall2,adelberger}
\begin{equation}
S_{\rm eff-S}(E) = S(E_0) \left\{ 1 +\frac{1}{\tau} \left[\frac{5}{12} +
\frac{5 S'(E_0) E_0}{S(E_0)} + \frac{S''(E_0)
E_0^2}{S(E_0)}\right]\right\}\;.
\label{eq-seff-a}  
\end{equation}
The approximation depends on $S$ and its first two derivatives, all
evaluated at $E=E_0$. By expanding $S$ about $E=0$~MeV we obtain an
expression where we only need $S$ and its derivatives at the
origin. This results in the approximation \cite{adelberger}
\begin{equation}
S_{\rm eff-T} = S(0) \left[ 1 + \frac{5}{12 \tau} + \frac{S'(0) (E_0
+ \frac{35}{36} kT)}{S(0)} + \frac{S''(0) E_0}{S(0)}
\left(\frac{E_0}{2} + \frac{89}{72} kT\right)\right].
\label{eq-seff-b}
\end{equation}
Eqs.~(\ref{eq-seff-a}) and (\ref{eq-seff-b}) both explicitly depend on
the first and second derivatives of the $S$ factor. Another
approximation may be made in which the derivatives in
Eq.~(\ref{eq-seff-a}) are neglected, giving a much simpler form:
\begin{equation} 
S_{\rm eff-Snd} = S(E_0) \left\{ 1 +\frac{5}{12 \tau} \right\}.
\label{eq-seff-c}
\end{equation}
Finally we may replace $S(E_0)$ with $S(20)$ in this last equation to
yield
\begin{equation} 
S_{\rm eff-S20} = S(20) \left\{ 1 +\frac{5}{12 \tau} \right\}.
\label{eq-seff-d}
\end{equation}

To evaluate these approximations we need a functional form for
$S(E)$. For $E < 100$~keV, the functional form of $S$ for the
$^7$Be($p,\gamma$) reaction is well approximated by \cite{jennings}
\begin{equation}
S(E)/S(0) = \frac{0.0409}{0.1375 + E} + 0.703 + 0.343 E
\label{eq-pade}
\end{equation}
where $E$ is in MeV. With this expression for $S(E)/S(0)$ and taking
$S(0) = 19.0$~eVb \cite{jennings}, we have calculated $S_{\rm eff}$
and the different approximations to it.  The results are given in
Table~\ref{table-error} for a range of temperatures that includes
those relevant to the production of $^8$B in the sun, i.e.~$1.3 \times
10^7~{\mathrm K} < T < 1.6 \times 10^7$~K \cite{adelberger}. The
values of $S_{\rm eff}$ calculated with any of the approximations
deviate from the exact results, Eq.~(\ref{eq-seff}), by no more than
0.1~eVb or 0.5\%. Also, there is less than a 0.1~eVb variation in
$S_{\rm eff}$ over the given range of temperatures. We note
particularly the accuracy of Eq.~(\ref{eq-seff-d}), which requires
only the value of $S$ at 20~keV, and does not require any explicit
knowledge of the derivatives or of the temperature dependence in
$E_0$.

Since the first and second derivatives of the $S$ factor enter into
two of the approximations, an understanding of the accuracy with which
these derivatives can be determined is useful. In
Table~\ref{table-derivatives}, we show the derivatives given in
Ref.~\cite{jennings} for a range of models. The first four are hard
sphere models \cite{jennings} depending on the given hard sphere
radius, $r_c$. The next three are Woods-Saxon models.  These models
agree with generator coordinate model calculations for low energies,
$E < 300$~keV \cite{jennings}, so the conclusions drawn here will also
apply to results from the generator coordinate model calculations.

The first derivative shows more model dependence than the second
derivative. However even for the first derivative the model dependence
is quite small. Neglecting the hard sphere model with $r_c=4.1$~fm,
which was shown to be unrealistic \cite{jennings}, and the results
from Adelberger {\em et al.} \cite{adelberger}, the average of the
values of the first derivative is $(1/S) (dS/dE)|_{E=0} =
-1.87\pm0.1$~MeV$^{-1}$. This introduces a variation of 0.2\% in
$S_{\rm eff}$ and is negligible.

There is surprisingly little variation in the second derivative. In
the hard sphere model it does not depend on the radius over the range
shown. The results calculated using the Woods-Saxon models show only a
slight variation. Part of this variation may be due to the difficulty
in determining the second derivative from the numerical calculations. We
recommend $[1/(2S)](d^2S/dE^2)|_{E=0} = 15.7$~MeV$^{-2}$; the
uncertainty in the second derivative introduces a negligible error in
$S_{\rm eff}$.

There has been considerable disagreement in the literature on what
values should be used for the derivatives. (See for example,
Ref.~\cite{barker}). The last line of Table~\ref{table-derivatives}
gives values of the derivatives from Adelberger {\em et al.}
\cite{adelberger} that are very different from those we have obtained
in any of the other models.  However, we can reproduce their results
by following their procedure for determining the derivatives and
fitting $S(E)$ by a quadratic form over the range $E =
20$ -- 300~keV. Similarly, we can reproduce the values Barker obtained
\cite{barker} for the derivatives by fitting a quadratic form over
the range he chose, 0 -- 100~keV. An accurate determination of the
derivatives from a fit of a quadratic form could only be obtained if
the fit region was restricted to $E<10$~keV.  Thus the differences
in the values of derivatives quoted in the literature are due to the
range of energies used for the fits and are unrelated to the true
model dependence of the derivatives at threshold.

We show in Table~\ref{table-error2} the results of using
Eq.~(\ref{eq-seff-b}) with two different sets of derivatives. The
first is that presented above, namely $-1.87$~MeV$^{-1}$ and
15.7~MeV$^{-2}$, and the second is that from Adelberger {\em et al.}
\cite{adelberger} given in Table~\ref{table-derivatives}. Using this
second set of derivatives introduces an error of 0.3~eVb in $S_{\rm
eff}$. As the reaction rate must be known to better than 5\%
\cite{adelberger}, this 1.5\% error is 30\% of the total error
allowed. Remarkably, using the inaccurate values of the
derivatives at threshold is worse than neglecting the derivatives
entirely and using $S(20)$.

In conclusion the use of approximate forms for $S_{\rm eff}$ does not
generate significant errors if accurate parametrisations of the $S$
factor and its derivatives are used. Eq.~(\ref{eq-seff-d}) is of
special interest since it avoids the use of the derivatives by using
$S$ evaluated at 20~keV. We therefore recommend that determinations of
the $S$ factor for $^7$Be($p,\gamma$)$^8$B at low energy quote $S(20)$
as well as $S(0)$.

\acknowledgements{This work was motivated by discussions with
C.W.~Johnson. Financial support from Natural Sciences and Engineering Research
Council of Canada is gratefully acknowledged.}

\begin{table}
\caption[]{The values of $S_{\rm eff}$, in eVb, from the different
approximations using Eq.~(\ref{eq-pade}) for $S(E)$. $E_0$ is in keV
and $T$ is in $10^6$~K.}
\label{table-error}
\begin{tabular}{ccccccc}
$T$ & $E_0$ & $S_{\rm eff}$ & $S_{\rm eff-S}$ &  $S_{\rm eff-T}$
 & $S_{\rm eff-Snd }$ &  $S_{\rm eff-S20}$  \\
\hline
12 & 15.4 & 18.67 & 18.68 & 18.69 & 18.70 & 18.58 \\
13 & 16.3 & 18.65 & 18.66 & 18.68 & 18.69 & 18.59 \\
14 & 17.1 & 18.64 & 18.65 & 18.66 & 18.67 & 18.59 \\
15 & 17.9 & 18.62 & 18.63 & 18.64 & 18.65 & 18.60 \\
16 & 18.7 & 18.60 & 18.61 & 18.63 & 18.64 & 18.60 \\
17 & 19.4 & 18.58 & 18.59 & 18.61 & 18.62 & 18.61 \\
\end{tabular}
\end{table}

\begin{table}
\caption[]{The first and second derivatives of $S$ evaluated at
threshold for the models of Ref.~\cite{jennings}. The nomenclature is
also that of Ref.~\protect\cite{jennings}.}
\label{table-derivatives}
\begin{tabular}{ccc}
Model & $ \displaystyle \frac{1}{S} \frac{dS}{dE}$ (MeV$^{-1}$) & 
$ \displaystyle \frac{1}{2S}\frac{d^2S}{dE^2}$
(MeV$^{-2}\frac{\strut}{\strut}$)  \\\hline   
$r_c = 0.0$~fm & $-$1.77 & 15.7 \\
$r_c = 1.0$~fm & $-$1.82 & 15.7\\ 
$r_c = 2.4$~fm & $-$1.92 & 15.7 \\ 
$r_c = 4.1$~fm & $-$2.09 & 15.7 \\ 
B1             & $-$1.91 & 16.1 \\
B2             & $-$1.84 & 16.0 \\
T              & $-$1.96 & 15.7 \\
Adelberger {\em et al.} \cite{adelberger}
               & $-$0.70 & \phantom{0}1.9  \\ 
\end{tabular}
\end{table}

\begin{table}
\caption[]{$S_{\rm eff-T}$, in eVb, as obtained using the two different sets of
derivatives discussed in the text. The temperature is in $10^6$~K.}
\label{table-error2}
\begin{tabular}{lcccccc}
$T$ & 12 & 13& 14& 15& 16& 17\\\hline
$S_{\rm eff-T}$ (eVb) & 18.68&18.66&18.64&18.62&18.61&18.59\\
$S_{\rm eff-T}$ (eVb) & 18.97&18.96&18.96&18.95&19.94&18.94
\end{tabular}
\end{table} 

\end{document}